\documentstyle[mprocl]{article}

\input{psfig}
\def\Journal#1#2#3#4{{#1} {\bf #2}, #3 (#4)}

\def\PRB{{\em Phys. Rev.} B}
\def\PC{{\em Physica} C}
\def\RMP{{\em Rev. Mod. Phys.} }
\def\ZPB{{\em Z. Phys.} B}

\def\vep{\varepsilon}

\def\be{\begin{equation}}
\def\ee{\end{equation}}
\def\bea{\begin{eqnarray}}
\def\eea{\end{eqnarray}}

\begin{document}
\title{DYNAMICAL PAIRING CORRELATIONS IN THE t--J MODEL 
WITH NON--ADIABATIC HOLE--PHONON COUPLING}
\author{G. WELLEIN, B. B\"AUML, and H. FEHSKE}
\address{Physikalisches Institut, Universit\"at Bayreuth,
\\ D--95440 Bayreuth, Germany
\\E-mail: holger.fehske@theo.phy.uni-bayreuth.de}   
\maketitle\abstracts{
We examine the effects of hole--phonon interaction on the 
formation of hole pairs in the 
2D Holstein t--J model. 
Using finite--lattice diagonalization techniques,
we present exact results for the two--hole binding energy 
and the s-- and d--wave pairing  susceptibilities.
} 
The interplay of electronic and lattice degrees of freedom 
in strongly electron--correlated systems is now attracting a lot of
attention. This interest is partially due to the prominent role 
of the electron--phonon (EP) interaction in several transition metal
oxides with strong Coulomb correlations, such as the colossal 
magnetoresistive manganites or the charge--ordered (insulating) nickelates. 
For the high--$T_c$ superconducting cuprates, very 
recent experiments demonstrate the relevance of the coupling 
between the charge carriers and the lattice dynamics as well~\cite{ZHKM97}.

For the sake of simplicity, we describe the basic interactions 
in the latter systems by an effective single--band Hamiltonian, 
the so--called Holstein t--J model~\cite{WRF96} 
\begin{equation}
{\cal H}= {\cal H}_{t-J}^{}
- \sqrt{\varepsilon_p\hbar\omega_0}  \sum_i \big(b_i^{\dagger} +
b_i^{} \big) \tilde{h}_i^{}
+ \hbar\omega_0 \sum_i \big(b_i^{\dagger}b_i^{} +
\mbox{\small $\frac{1}{2}$}\big)\,, 
\end{equation}
which contains besides nearest--neighbour hole  
transfer ($t$) and antiferromagnetic (AFM) spin exchange ($J$)
on a square lattice, the coupling of doped holes  ($\tilde{h}_i$) 
to a dispersionsless optical phonon mode 
(representing, e.g., local apical--oxygen 
breathing vibrations). 
Here $\vep_p$ is the hole--phonon coupling strength
and $\hbar\omega_0$ denotes the phonon frequency. 
The single--particle excitations of the model (1) have been studied 
numerically~\cite{KMKF96,FWBB97}; 
the main result is that in the presence of strong AFM spin correlations 
even a moderate EP coupling can cause polaronic effects.

In this contribution, we focus on the two--hole subspace  
in order to comment on hole--pair formation. Employing the Lanczos
algorithm in combination with a well--controlled truncation
of the phononic Hilbert space~\cite{WRF96,FLW97}, 
we are able to calculate both ground--state and dynamical properties
preserving the full dynamics and quantum nature of phonons.
It is worth noticing, that our multi--mode treatment of the phonons differs 
significantly from the one--phonon calculation performed in Ref.~6
for the t--J model coupled to oxygen breathing or buckling modes.

The dynamical pair spectral function can be written as    
\begin{equation}
 {\cal A}_{2h}(\omega)= \sum_n |\langle {\mit \Psi}^{(N-2)}_n\,
|\,{\mit \Delta}_{\alpha}^{}\,|\, {\mit \Psi}^{(N)}_0\rangle |^2 
\;\,\delta\, [\omega -(E_n^{(N-2)}-E_0^{(N)})]\,,
\end{equation}
where ${\mit \Delta}_\alpha=\frac{1}{\sqrt{N}}\sum_{i;\delta=\pm x, \pm y}
  {\cal F}_{\alpha}(\delta)\tilde{c}_{i\uparrow}
\tilde{c}_{i+\delta\downarrow}$, with 
${\cal F}_{\alpha}(\pm x)=1$ and  ${\cal F}_{\alpha}(\pm y)= -1$ $[+1]$  
for d--wave [s--wave] pairing.
\begin{figure}[p]
\mbox{\psfig{figure=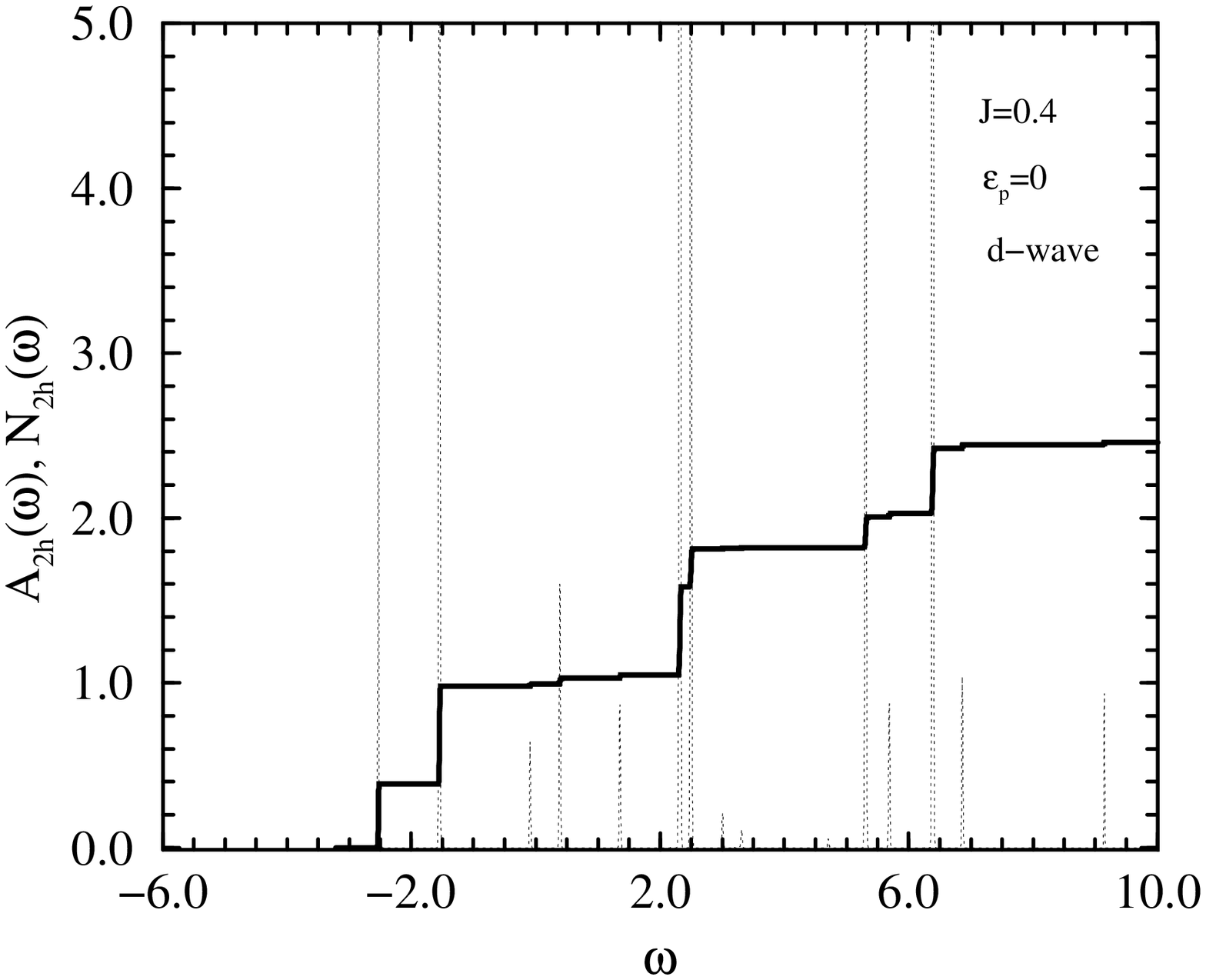,height=2.1in}\hspace*{4pt}
\psfig{figure=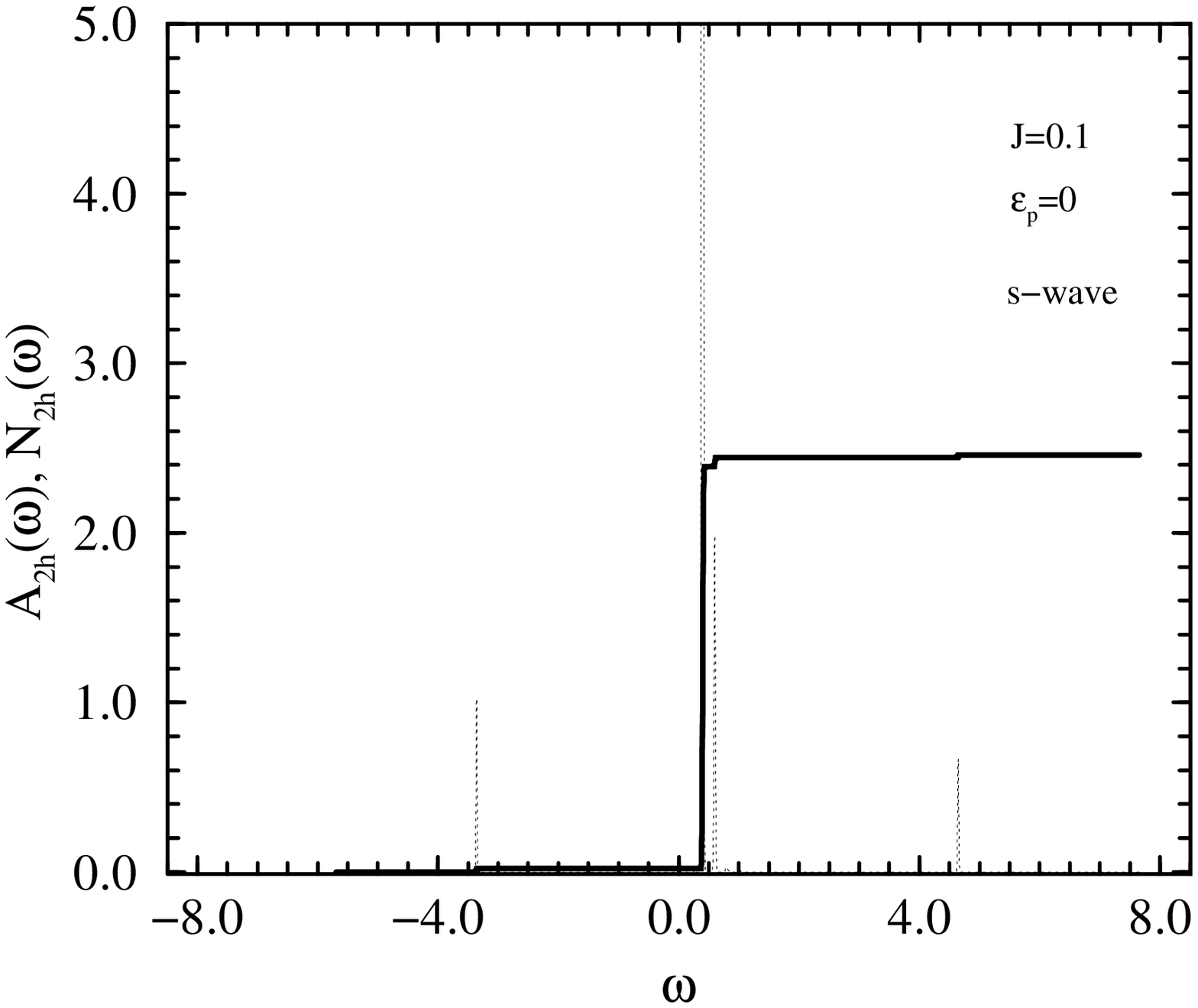,height=2.1in}}
\mbox{\psfig{figure=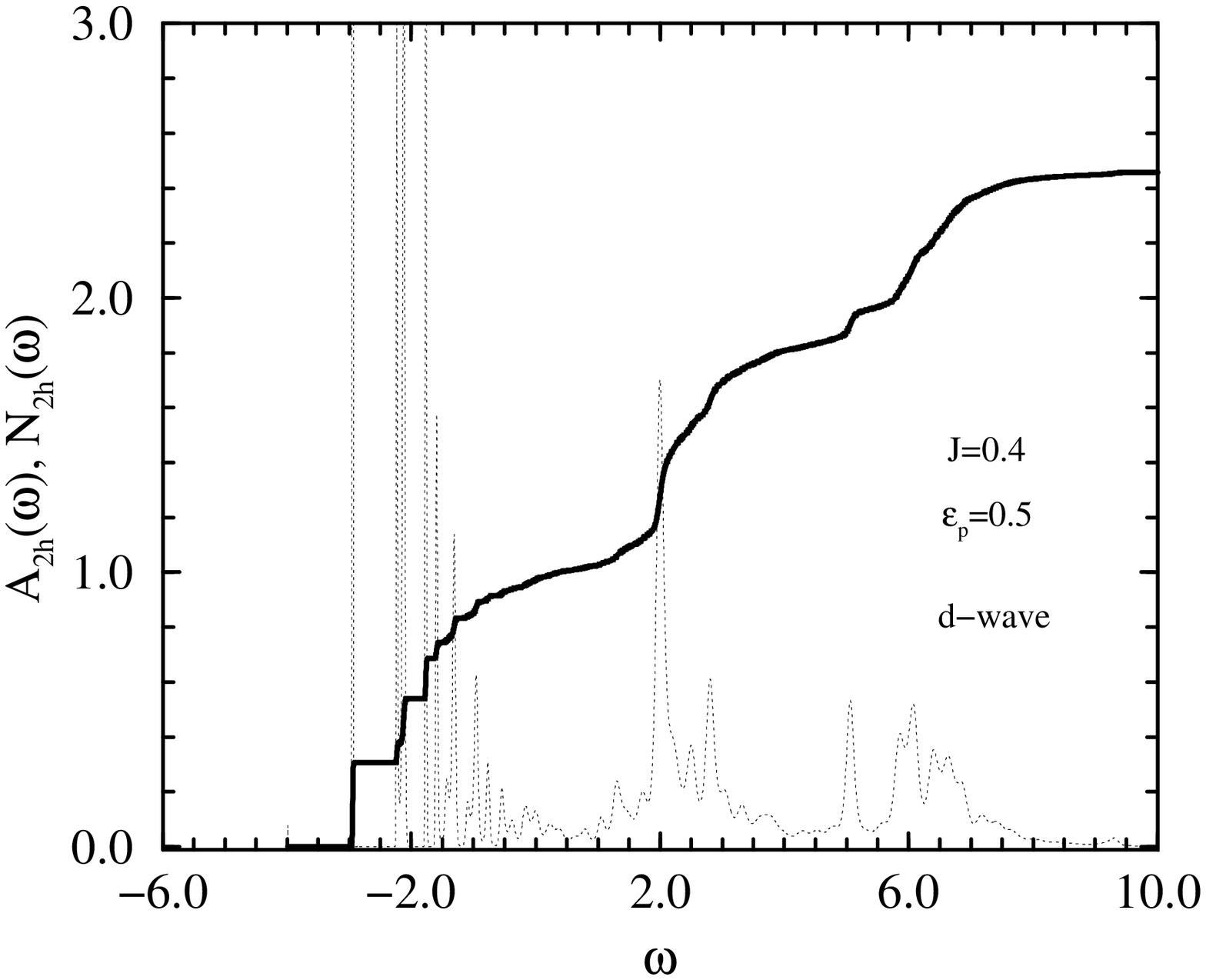,height=2.1in}\hspace*{4pt}
\psfig{figure=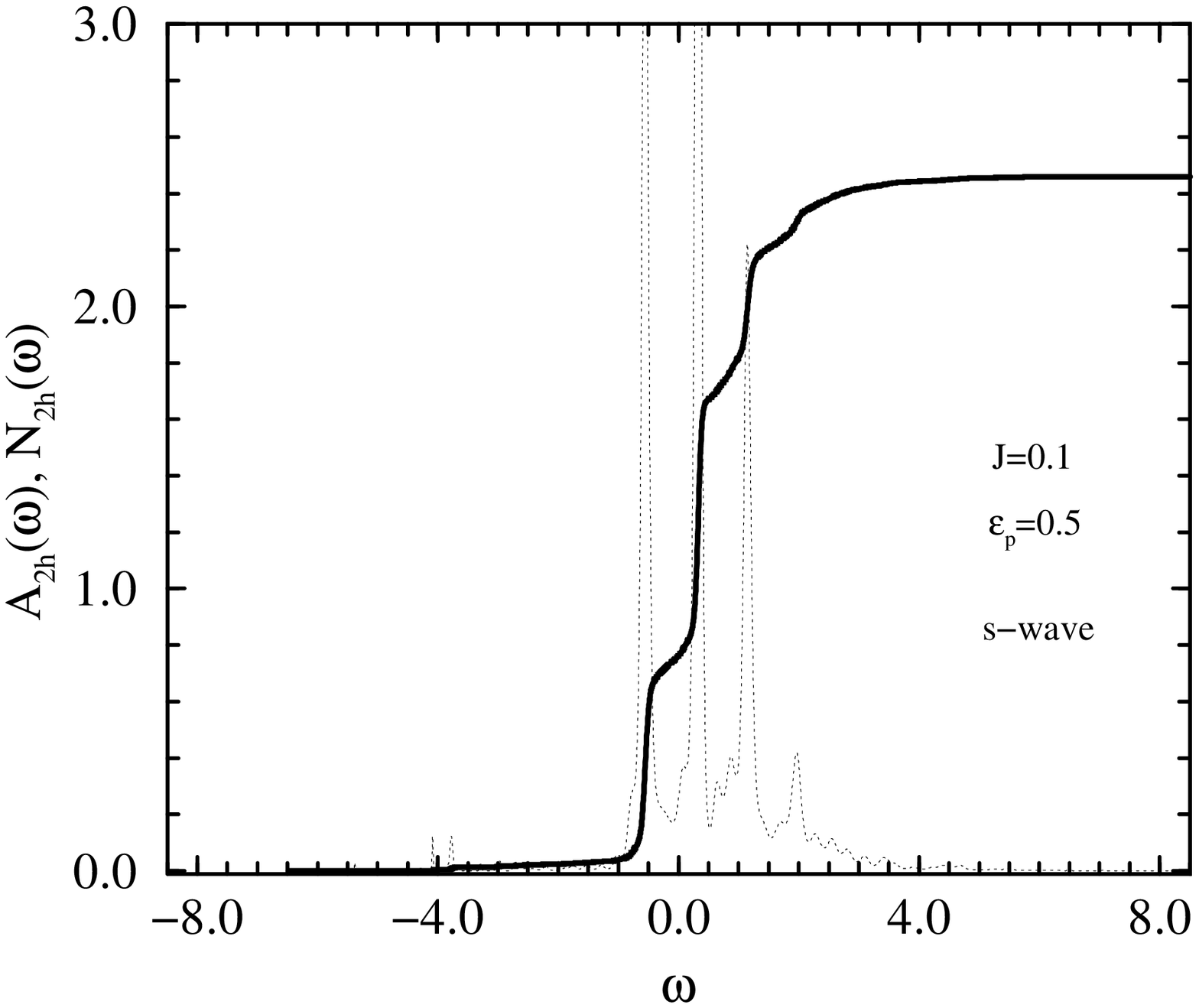,height=2.1in}}
\mbox{\psfig{figure=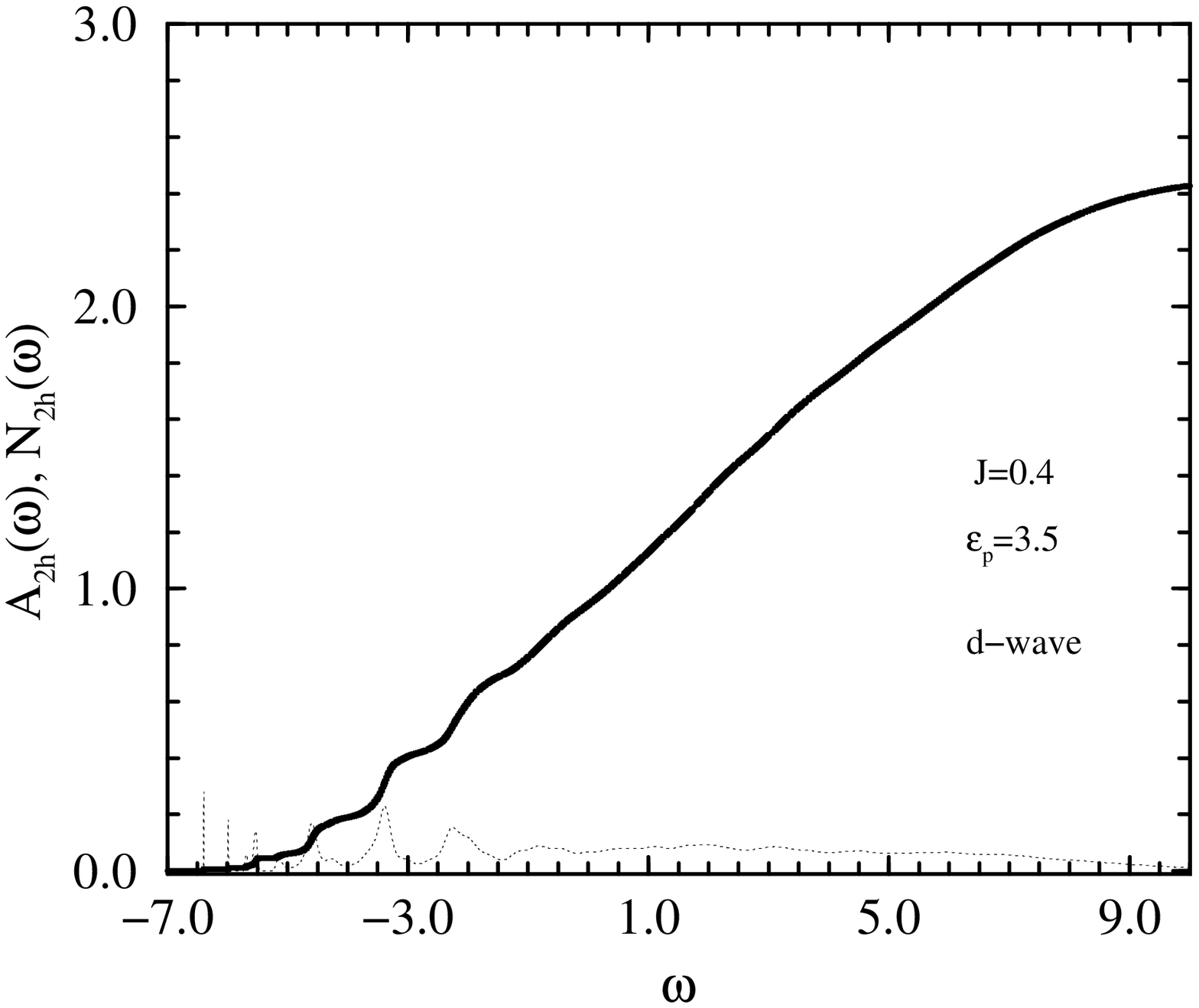,height=2.1in}\hspace*{4pt}
\psfig{figure=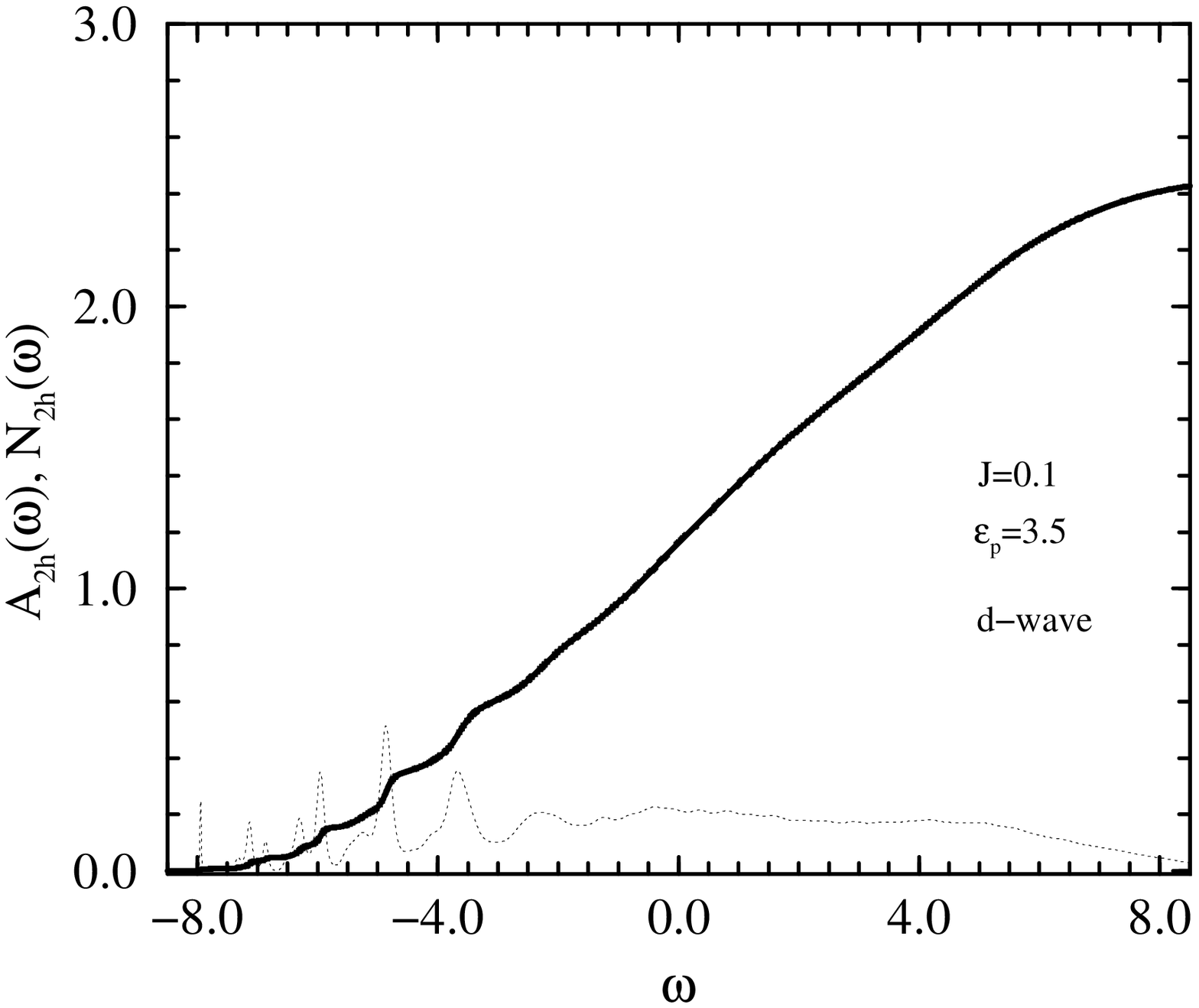,height=2.1in}}
\caption{Dynamical pair spectral function ${\cal A}_{2h}(\omega)$ 
(dotted lines) and integrated spectral weight  ${\cal N}_{2h}(\omega)= 
\int_{-\infty}^{\omega} d\omega^{\prime} \, {\cal A}_{2h}(\omega^{\prime})$
(bold lines) calculated for the 2D Holstein t--J model on a 
ten--site square lattice with periodic boundary conditions 
at $\hbar\omega_0 =0.8$.  
Depending on the model parameters $J$ and  $\vep_p$,  
results are presented only for those hole--pair wave 
functions $\,|{\mit \Delta}_{\alpha}\,
{\mit\Psi}_0^{(N)}\rangle$ which have a finite overlap 
with  $|{\mit\Psi}_0^{(N-2)}\rangle$.  Note that for the the pure t--J model
the symmetry of the  two--hole  ($\vec{K}=0$) ground state is changed
from d--wave $(B/C_4)$ to s--wave  $(A/C_4)$ at $J_c=0.2001$ ($N=10)$. 
On the other hand, the (symmetry) change in the spectra, observed   
by comparing the plots for $\vep_p=0.5$ and 3.5 at $J=0.1$, 
is driven by the EP coupling ($\vep_{p,c}\simeq 3.3$). 
All energies are measured in units of $t$. 
\label{fig:pairw08}}
\end{figure}

Numerical results for the spectral functions of the pairing operators 
${\mit \Delta}_\alpha$ are shown in Fig.~1 for exchange 
interactions $J=0.4$ (left column) and $J=0.1$ (right column) 
corresponding to two different regimes in the pure t--J model 
(see upper panels). For $J=0.4$, the d--wave pair spectrum 
exhibits a well--separated low--energy peak containing an appreciable amount 
of spectral weight which grows if $J$ is enhanced (cf. the inset of
Fig. 2). Since the rest of the spectrum becomes incoherent 
with increasing lattice size $N$, the dominant peak at the bottom of 
the spectrum has been taken as signature of a d-wave quasiparticle 
bound state~\cite{Da94}. By contrast, the s--wave spectrum shows no 
such quasiparticle--like excitation. 
In the weak EP coupling case, the main features of the $\vep_p=0$ 
spectra are preserved, although, of course, additional 
phonon satellite structures appear (cf. the discussion of Fig. 3 below). 
The situation is drastically different in the strong--coupling
regime. Here a strong mixing of electron and phonon degrees of freedom
takes place and less mobile (bi)polaronic charge carriers emerge.
The polaronic self--trapping transition is accompanied by a 
dramatic reduction of the coherent band width and, as a result,
the AFM spin interaction becomes much more effective.
Therefore, at  $\vep_p=3.5$, the spectrum for $J=0.1$   
looks very similar to that for $J=0.4$.

The relative spectral weight,  ${\cal Z}_{2h}$,
located in the lowest pole of ${\cal A}_{2h}$, 
is plotted in Fig. 2 (left panel). Obviously, we observe  
a strong suppression of the d--wave quasiparticle
residue with increasing EP coupling. 
However, ${\cal Z}_{2h}$ gives only a measure of 
the ``electronic'' contribution to the d--wave bound state.
In fact, according to previous work~\cite{SPS97}, 
composite pair operators $\bar{\mit \Delta}_{\alpha}$, 
properly dressed by a phonon cloud, give large quasiparticle 
weights in $\bar{\cal A}_{2h}$.
That the Holstein EP interaction may stabilize a bound state of 
two holes is clearly demonstrated by the behaviour of the 
binding energy $E_B^2$, which has been calculated for the 
larger $4\times 4$ lattice in order to reduce finite--size effects
(see the right panel of Fig.~2). Whereas the hole attraction 
($E_B^2<0$) is less affected in the anti--adiabatic limit, 
the EP coupling promotes the pairing correlations 
between two holes in the adiabatic regime due to 
subtle retardation effects.   
This is to be contrasted with the findings 
for a coupling of the holes to the in--plane oxygen breathing mode which leads
to an hole repulsion for all frequencies and EP interaction  
strengths~\cite{SPS97}. 
\begin{figure}[h]
\vspace*{-.3cm}
\mbox{\psfig{figure=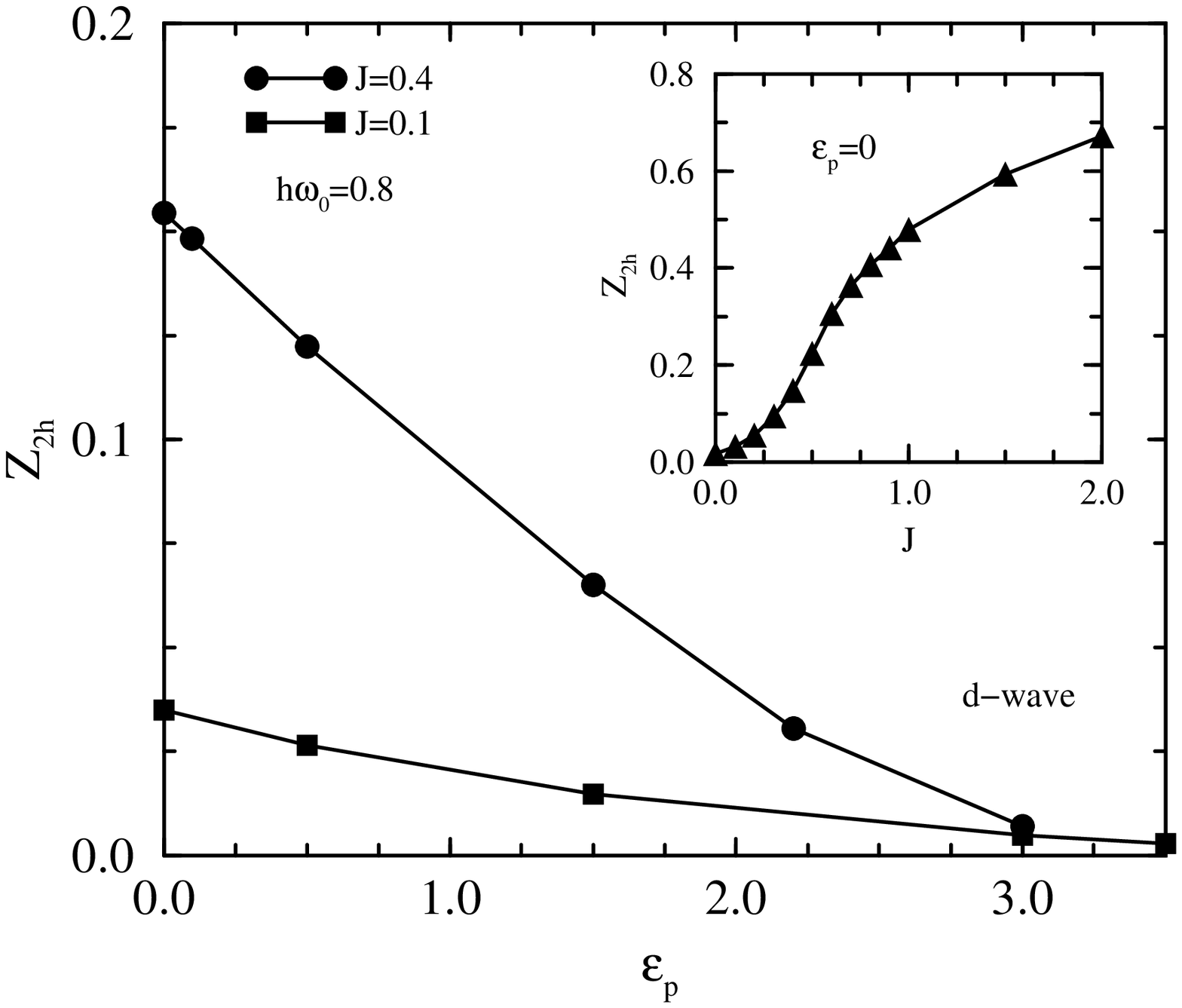,height=2.1in}\hspace*{4pt}
\psfig{figure=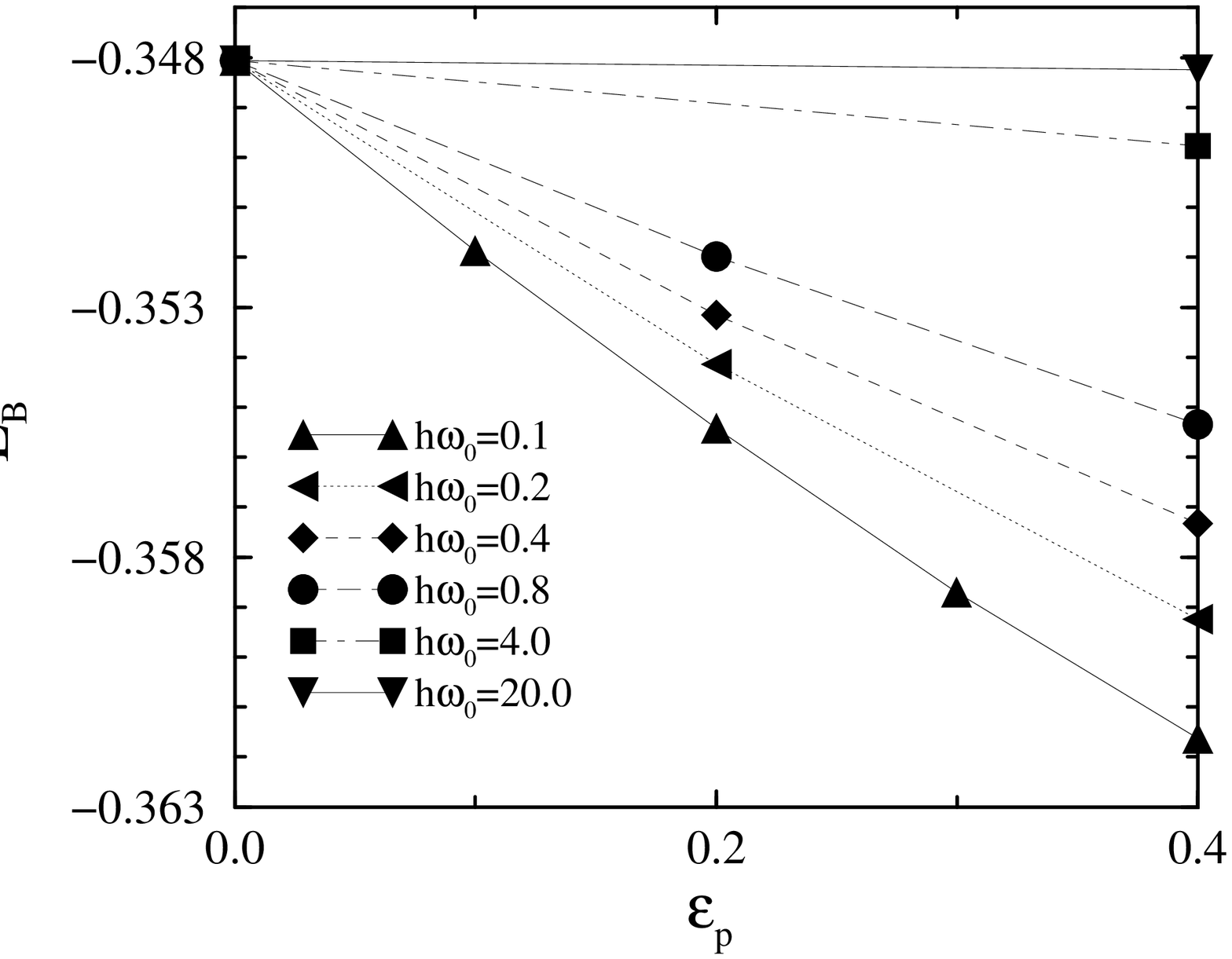,height=2.1in}}\vspace*{-.1cm}
\caption{Two--hole ``quasiparticle weight'' ${\cal Z}_{2h}=
|\langle {\mit\Psi}_0^{(N-2)}\,|\,
{\mit \Delta}_{\alpha}\,|\,
{\mit\Psi}_0^{(N)}\rangle|^2/
|\langle {\mit\Psi}_0^{(N)}\,|{\mit \Delta}_{\alpha}^{\dagger}
{\mit \Delta}_{\alpha}^{}|\,
{\mit\Psi}_0^{(N)}\rangle|$ (left panel; $N=10$)
and hole ``binding energy'' $E_B^2=E^{(N-2)}_0+E^{(N)}_0-2E^{(N-1)}_0$ 
(right panel; $N=16$ ) shown as a function of EP coupling strength $\vep_p$ at 
various phonon frequencies $\hbar\omega_0$. 
\label{fig:z2h_eb2}}
\end{figure}

Let us discuss the weak--coupling case in more detail. 
Fig.~3 presents the spectral decomposition of
the d-wave pairing operator at $\hbar\omega_0=0.1$ and 3.0. 
In the high phonon frequency (anti--adiabatic) regime, 
the pairing susceptibility behaves qualitatively in a similar way 
as the $\vep_p=0$ limit, in particular the low--energy part 
of the spectrum is given by purely electronic resonances.
\begin{figure}[h]\vspace*{-.4cm}
\mbox{\psfig{figure=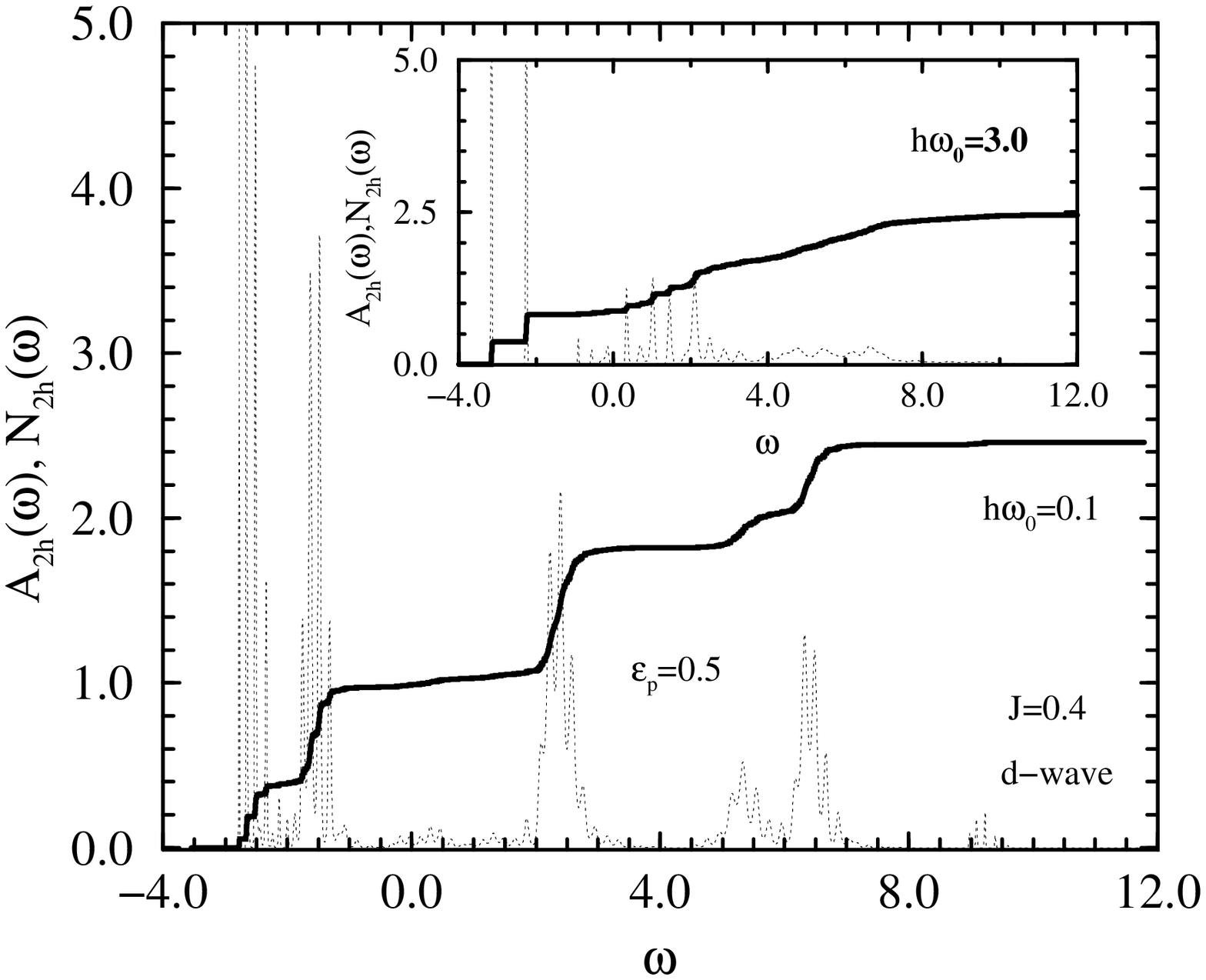,height=2.1in}\hspace*{4pt}
\psfig{figure=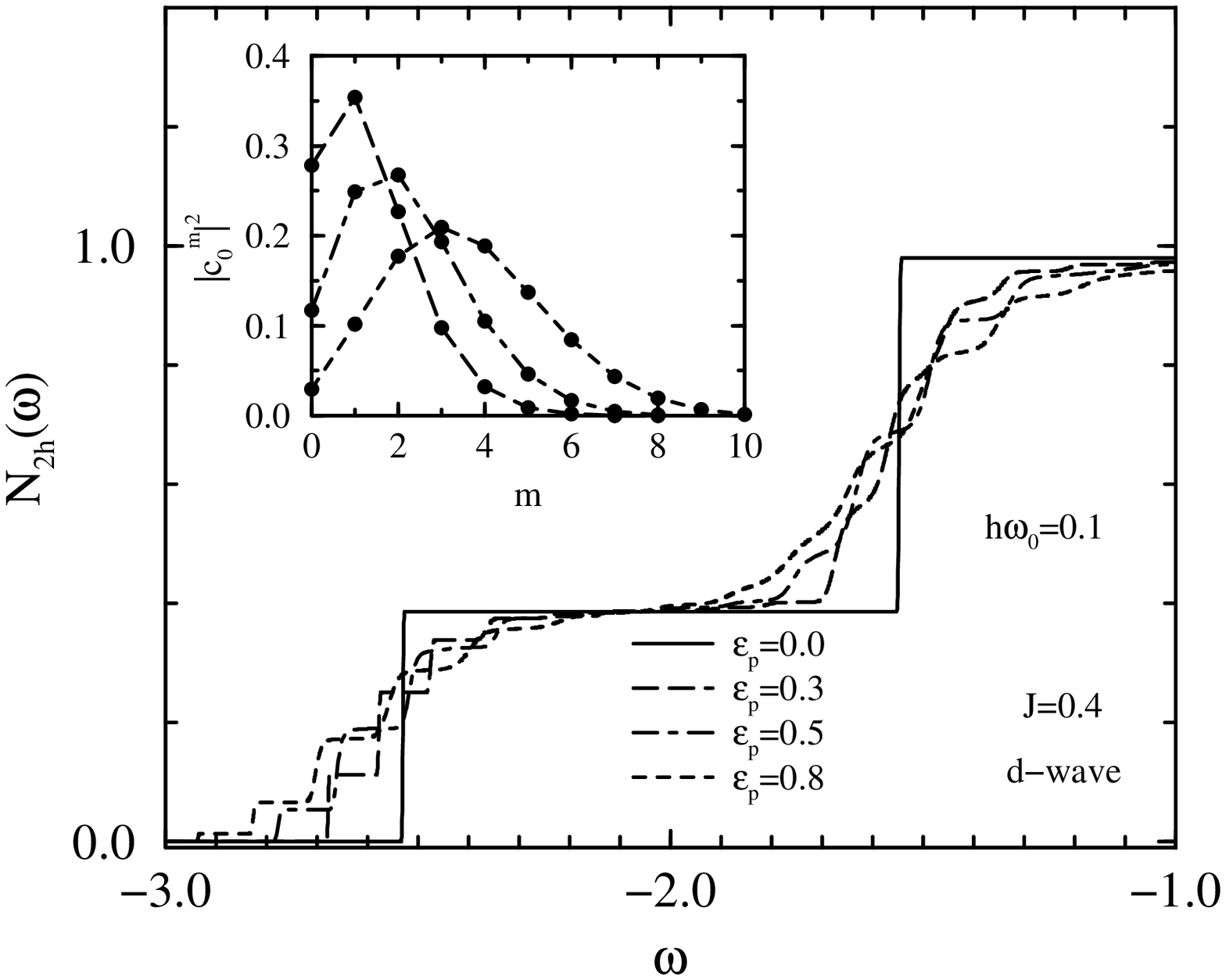,height=2.1in}}\vspace*{-.3cm}
\caption{Left panel: d--wave pair spectra ${\cal A}_{2h}$ at $\vep_p=0.5$ and 
$\hbar\omega=0.1$ (inset: $\hbar\omega=3.0$). Right panel: 
integrated spectral weight ${\cal N}_{2h}(\omega)$ in the low--energy 
part of ${\cal A}_{2h}$  for various $\vep_p$  
(inset: phonon--weight function $|c^m_0|^2$ in the two--hole 
ground state for the same parameters) \label{z2hw08}}
\end{figure} 
If the phonon frequency is much smaller than the electronic gaps, 
we found series of predominantly phononic side bands, being separated by  
$\hbar\omega_0$ and roughly centered around 
the positions of the electronic excitations.
The relative weights of these $\delta$--like peaks 
can be deduced from the corresponding jumps in ${\cal N}_{2h}(\omega)$
depicted for different $\vep_p$ in the right panel of Fig.~3.
Focusing on the lowest set of phonon sub-bands, it is interesting to note,    
that the weight of the {\it zero--phonon state} ($|c^0_n|^2$)
in the first {\it excited states} ($|{\mit \Psi}_n^{(N-2)}\rangle,\;
n=1,\,2,\ldots$), which is measured by ${\cal N}_{2h}(\omega)$, 
is approximately the same as the weight of the  
$m$--{\it phonon states} ($|c^m_0|^2$)  
in the {\it ground state} ($|{\mit \Psi}_0^{(N)}\rangle$).  
The definition of the coefficients $|c^m_0|^2$ is given in Refs.~4,5.  
A qualitative understanding of this fact can be obtained from 
the study of the independent boson model~\cite{Ma90},
where we can show exactly that  $|c^m_0|^2=|c^0_m|^2$ holds~\cite{We98}.

To summarize, our exact diagonalization studies of the 
2D Holstein t--J model give evidence for significant EP coupling effects. 
Most notably, the hole pairing in the t--J model may 
be stabilized by a dynamical (Holstein) hole--phonon interaction.   

\eject
\end{document}